# Quantized grain boundary states promote nanoparticle alignment during imperfect oriented attachment

*Andrew P. Lange, Amit Samanta, Tammy Y. Olson, Selim Elhadj[*]*

[*]*Corresponding author*

Lawrence Livermore National Laboratory

Mail-stop 470, 7000 East Ave. Livermore, CA 94550

**ABSTRACT**  Oriented attachment (OA) has become a well-recognized mechanism for the growth of metal, ceramic, and biomineral crystals. While many computational and experimental studies of OA have shown that particles can attach with some misorientation then rotate to remove adjoining grain boundaries, the underlying atomistic pathways for this "Imperfect OA" process remain the subject of debate. In this study, molecular dynamics and in situ TEM were used to probe the crystallographic evolution of up to 30 gold and copper nanoparticles during aggregation.  It was found that Imperfect OA occurs because (1) grain boundaries become quantized when their size is comparable to the separation between constituent dislocations and (2) kinetic barriers associated with the glide of grain boundary dislocations are small. In support of these findings, TEM experiments show the formation of a single crystal aggregate after annealing 9 initially misoriented, agglomerated particles with evidence of dislocation slip and twin formation during particle/grain alignment. These observations motivate future work on assembled nanocrystals with tailored defects and call for a revision of Read-Shockley models for grain boundary energies in nanocrystalline materials.

The aggregation of small crystalline particles occurs ubiquitously in natural geologic environments [1,2] and during industrial processes such as powder synthesis, sintering, and catalysis [3-5]. The rotation and alignment of individual particles during aggregation to form larger single crystal units, or grains, has been extensively observed and is often referred to as oriented attachment (OA), oriented aggregation, or crystallization by particle attachment (CPA) [6]. While the driving force for OA is incontrovertible as GBs have higher internal energies than perfect crystals, a consensus has not been reached regarding how, mechanistically, particles reorient themselves. In solution, it is often assumed that particles reorient themselves prior to or during attachment via van der Waals, electrostatic, and hydrodynamic interactions [2, 7]. However, it has also been hypothesized that particles may attach with some misorientation followed by the evolution of their microstructure to form a perfect single crystal [2, 6, 8, 9]. This so-called "Imperfect OA" [8] process is particularly interesting because it not only captures one pathway by which large single crystals may form from many small nuclei in solution (the prototypical problem of OA) but is also relevant to how nanoparticles orient themselves on crystalline substrates (e.g. during epitaxy) and grain growth during the sintering of nanocrystalline (NC) compacts.

In principle, multiple atomistic mechanisms exist by which GBs can be removed after attachment during Imperfect OA. These mechanisms can be sorted into a few categories involving (i) the diffusion of 0-dimensional objects (i.e. diffusional flow), (ii) dislocation slip processes, or (iii) grain boundary migration as described below.

(i) The diffusion of 0-dimensional objects (e.g. atoms or point defects such as interstitials) may facilitate recrystallization of a material. Some examples include Ostwald ripening during which atoms from a small particle move to the surface of a larger particle due to surface energetics, GB restructuring during particle rotation (i.e. atomic diffusion within a GB), and dislocation climb which necessitates the motion of point defects.

(ii) Dislocation slip can cause particles or grains to deform thereby decreasing the misorientation of their GB.

(iii) Grain boundary migration may occur in which a 2-dimensional GB plane moves through a particle/grain toward a nearby surface. This can be facilitated by either atomic diffusion (i) or dislocation slip (ii) and may therefore not require its own category. However, it has been differentiated here because it captures a separate, well established mechanism.

Grain alignment in bulk polycrystalline materials with micrometer-sized grains is generally thought to occur by this last mechanism (iii), capillary-driven grain boundary (GB) migration [10,11]. Associated models have been widely successful at predicting grain growth and microstructural evolution during thermal processing without a priori knowledge of GB structure or the atomistic details of GB motion [12,13]. In contrast, during Imperfect OA, nanograin coarsening (or particle alignment) has been found to occur by the rotation of individual particles or grains resulting in "coalescence" when the crystallographic axes of adjacent grains align [11-21]. This can only occur by (i) or (ii). In the context of nanograin growth in bulk NC materials (e.g. during sintering), phenomenological models have assumed that grain rotation occurs by (i), the diffusion of atoms within the GB to accommodate changes in GB structure [25-28]. Further, these models assume the driving force for rotation follows the continuum Read-Shockley model for the change in GB energy [22-25]. However, both experimental in-situ transmission electron microscopy (TEM) studies and molecular dynamics (MD) simulations have found that nanograin rotation is accompanied by dislocation slip, consistent with mechanism (ii) [19,22,29-37].

In this study, MD simulations and in-situ TEM experiments involving up to 30 Au and Cu particles were carried out to examine particle alignment during aggregation. This is the first in silico MD investigation to the authors' knowledge of grain evolution in freely sintered agglomerates with this many particles. Observations presented here suggest grain boundaries are removed by dislocation activity (ii) which is consistent with previous observations involving 2-particle coalescence [29]. Grain boundary energy and potential energy barrier calculations were carried out to examine this mechanism in detail. It was found that an instability arises when the size of a GB becomes comparable to the distance between constituent dislocations. This gives rise to a "quantization" of stable GB dislocation structures and promotes rapid grain alignment during the initial stages of aggregation when GB dislocations have low kinetic barriers to glide to nearby surfaces. This is yet another example of size quantization when dimensions are shrunk to the nanoscale and provides an important conceptual framework for models and experiments involving the aggregation of nanoparticles. Further, these results also call for a revision of Reed-Shockley models for GB energies in NC materials to adequately capture grain growth and crystal growth kinetics.

Molecular dynamics simulations were carried out at elevated temperatures (~830 °C) to accelerate the kinetics of Imperfect OA and provide a generalized picture of the associated phenomena. Nanoparticles were initiated with either a spherical or Wulff shape and had randomized diameters ranging from 3 to 20 nm. They consisted of either Cu or Au atoms. Figure 1 provides representative snapshots from one simulation involving 10 Au particles to illustrate the observed behavior. As shown in Fig. 1(a), the 10 particles were misoriented at the beginning of the simulation but evolved to form a single crystal with ~7 twin boundaries (TBs) within 30 nanoseconds (ns) (note, not all TBs are visible from the orientation shown). In all simulations, particles underwent rapid rotation during neck formation as the atoms on adjacent facets began to interact. As shown in the supplemental movie S1, dislocations formed along interfaces between misaligned particles during attachment to form stable GBs. Grain boundary dislocations then gradually glided to nearby surfaces over longer timescales. This is evident in Fig. 1(b) which shows a dramatic decrease in the number of non-twin grain boundaries (NTBs) within the first 1 ns of the 10-particle simulation shown in Fig. 1(a) (see Fig. S2 for complementary plots for other simulations). Many larger NTBs, however, remained for up to 30 ns. There was a corresponding increase in the number of twin boundaries (TBs) in all simulations involving more than 3 particles due to both twin nucleation by surface shearing and the evolution of many NTBs to stable TBs. This was followed by the slow, incomplete removal of TBs via the motion of Shockley partial dislocations. Generally, the GBs between particles that were free to rotate along the perimeters of aggregates were removed quickly, while those formed near the interior between more than two particles remained for longer.

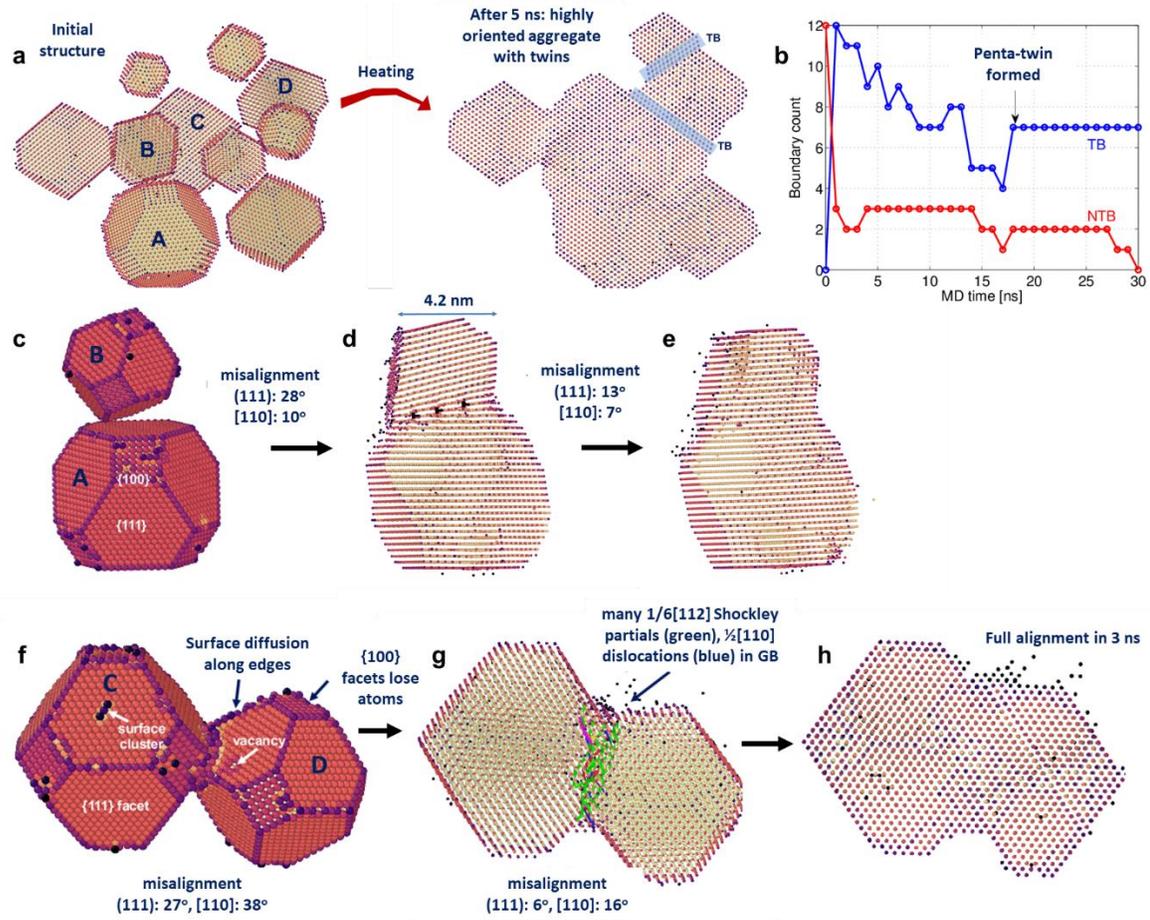

**Figure 1**. Snapshots of MD simulation in which 10 randomly oriented, Wulff-shaped nanoparticles were annealed for ~40 ns at ~830 C. (a) Single crystal aggregate formed with remnant twin boundaries (see movie S1). (b) Rapid decrease in number of NTBs during first 1 ns; corresponding increase, then slow decrease in number of TBs. (c)-(e) Alignment of particles marked A and B in (a) with 28º initial misalignment between nearest (111) surface facets. (f)-(h) Alignment/coalescence of particles marked C and D in (a).

This latter observation highlights how the number of particles in an aggregate can influence single crystal formation during Imperfect OA. When individual particles attach to the outside of an aggregate (or particle), they can rotate in any direction to accommodate GB removal. On the other hand, if particles are attached on multiple sides (e.g. in the interior of an agglomerate), they become pinned which creates a stress that opposes dislocation motion to the surface. For example, while all NTBs were removed in the simulation involving 10 particles (Fig. 1b), some persisted in agglomerates involving 25 and 30 particles (Fig. S2) even after 40 ns. In contradiction to this, however, metastable GB structures (GBs not removed over timescales accessible to MD) were also observed in small aggregates involving only a few particles. In one simulation involving 3 particles (the minimum number of particles for which rotation may be restricted), a stable $\Sigma_3 + \Sigma_3 + \Sigma_9$ boundary was observed; similarly, a double penta-twin structure was observed after annealing a system initiated with 4 particles. These GB junctions were not removed even after 50 ns of simulation time at 830 ºC. Thus, the evolution of an aggregate toward a single crystal may be inhibited by NTBs near the interior of aggregates or the formation of GB junctions.

An illustration of the removal of a GB between two particles along the perimeter of the agglomerate in Fig. 1(a) is shown in Figs. 1(c)-(e). Although the initial misalignment between the (111) surface facets of particles A and B was large, ~28º, these two particles became aligned within ~150 ps. After making contact, local atomic shuffling and surface diffusion in the neck region aided the formation of edge dislocations at the particles' interface (Fig. S3). The lattice planes in the GB only contained a few atoms due to the small neck size. These atoms diffused rapidly over short distances to aid dislocation rearrangement which resulted in the structure shown in Fig. 1(d). The resulting three GB dislocations then glided on a set of {111} planes to the surface of the top particle (particle B) leaving a coherent TB at the particles' interface (note TB not visible in orientation shown in Fig. 1(e)).

Similarly, an illustration of the removal of a GB between two particles in the interior of the agglomerate in Fig. 1(a) is shown in Figs. 1(f)-(h). The GB between particles C and D grew after attachment. This was facilitated by the diffusion of surface atoms from high-energy {100} facets along {110} surfaces to the neck region as marked in Fig. 1(f). The resulting GB contained a complex array of full and partial dislocations of mixed edge and screw character (Fig. 1(g)) which gradually glided to nearby surfaces leaving behind a fully aligned aggregate after about 3 ns (Fig. 1(h)). This was followed by further shape evolution and densification mediated by surface diffusion. This sequence of events – full or partial particle alignment during attachment, additional alignment due to GB "disintegration" (i.e. the glide of GB dislocations to the surface), followed by shape evolution and densification – agrees with experimental observations of coalescence between 2 gold nanoparticles during in-situ TEM beam heating experiments as shown in movies S4 and S5. Note, snapshots from movie S5 also described in a previous publication [29].

The alignment of nanoparticles by the "disintegration" of GBs is unlike classical pathways for grain coarsening (e.g. Ostwald ripening or GB migration). To further examine this, potential energy barriers associated with the motion of dislocations from GBs to nearby surfaces were calculated using the nudged elastic band (NEB) method. The energy barrier for a dislocation to escape from a 10º symmetric tilt GB between two nearly spherical 10 nm nanoparticles was found to be ~1 eV as shown in Fig. 2(a). Therefore, if a GB dislocation is perturbed from a metastable position (one in which the stress forces are balanced), it can easily overcome lattice resistance and glide to nearby surfaces along {111} slip planes. The driving force for this may be image forces at the surface or neighboring GB dislocations. The glide of GB dislocations to nearby surfaces was previously found to be initiated by the nucleation of dislocations in the neck region between adjoining particles [29]. In agreement with this, NEB calculations showed that the barriers to shear small surface islands (monoatomic layer with ~5-10 atoms) are 0.40-0.70 eV for metals (Fig. S6), much smaller than the 5-10 eV barriers associated with the homogeneous nucleation of dislocations inside stress-free ~5 nm nanoparticles (Fig. S7). This suggests the surface of a nanoparticle is very dynamic (see movie S8) and thermally activated slip processes in the neck groove may destabilize GBs. This thermal surface shearing process led to the formation of many three-layer twins near the surfaces of small particles, evident in the rapid increase in TBs in Fig. 1(b) between 0 and 1 ns.

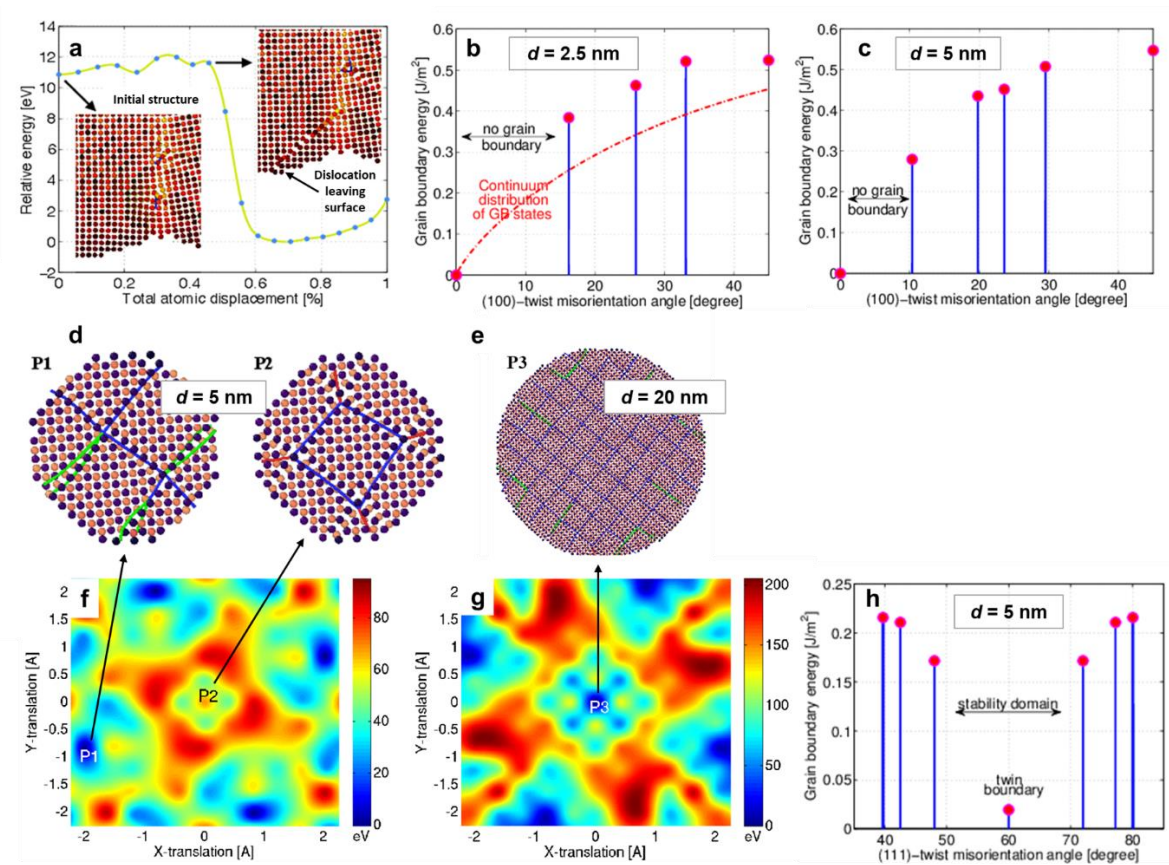

**Figure 2**. (a) Energy of 10° <100>-tilt GB between two nanoparticles (d=10 nm) as 1 of 4 edge-type dislocations glided to surface; energy of initial structure relative to final was 10 eV; barrier

to remove dislocation ~1 eV. (b)-(c) Number of accessible grain boundary states for <100>-twist boundary increases from 5 for d = 2.5 nm I (b) to 6 for d = 5 nm I (c). Continuum distribution of GB states using Read-Shockley model shown in red (b), taken from [39]. (d)-(g) Only 6 dislocation-segments exist in metastable 10° <100>-twist GB between two 5 nm particles (shown in (d), marked P1 in (f)). Number of dislocation segments increases by a factor of ~5 in GB between two 20 nm sized particles with the same misorientation (e). Number of GB states, with respect to how GB dislocations tile the GB, increases with nanoparticle size, evident from the number of local energy minima on the potential energy surfaces in (f) and (g) (d = 5 and 20, respectively). (h) Metastable states for <111>-twist boundary suggest GBs near this misorientation will "disintegrate" to form TBs.

The Read-Shockley model is commonly used to determine GB energies as a function of misorientation angle. This model assumes GB structures consist of arrays of dislocations with edge, screw, or mixed character depending on the misorientation plane and angle [38,39]. The energy of a GB in this model, $\gamma_\theta$, is a continuous function of the misorientation angle $\theta$: $\gamma_\theta = \gamma_o(A_o - \log\theta)$, where $\gamma_o$ and $A_o$ depend on the elastic constants of the material and the core energies of GB dislocations (red-curve in Fig. 2(b)) [39]. Energy profiles of [100] twist GBs in very small, $d$=2.5 and 5 nm, particles, however, are dramatically different from the continuum picture as shown in Fig. 2(b) and Fig. 2(c), respectively. A finite number of metastable states exist for $0°\leq\theta\leq 45°$ (5 in Fig. 2(b) and 6 in Fig. 2(c)), separated by large gaps in $\theta$ for which no metastable GB structures are accessible. This discreteness arises as only a few dislocations are present in a GB at the nanoscale for a given misorientation (e.g. $N=(2d/b)\sin(\theta/2)$) for symmetric tilt GBs, where b is the magnitude of Burgers vector. This discreteness is evident when comparing the GBs shown in Figs. 2(d) and (e) which both have twist misorientations of 10º but particle diameters of 5 and 20 nm, respectively. When GBs lose a dislocation and the number of dislocations goes from N to N-1, the misorientation jumps discontinuously from $2\sin^{-1}$ (N b/2d) to $2\sin^{-1}$ ((N − 1) b/2d), driven by residual elastic strain. The total energy of a GB, which depends on the length of the constituent dislocations, also exhibits discrete jumps with changes in θ (see Fig. S9). It is worth noting that finite element simulations have shown that a single dislocation can be stable at the boundary between two nanoparticles which may be interpreted as the first state with the lowest misorientation angle in Fig. 2(b) or (c) [40].

Nanoparticle GB states are not only discrete with respect to misorientation angle, but also with respect to how the dislocation network tiles the GB. For example, by translating one particle relative to the other less than the interatomic spacing in a direction within the GB plane, the associated GB dislocation network changes from the local minimum P1 to P2 for 5 nm particles with a twist misorientation of 10º as illustrated in Fig. 2(d). The energy of P1 is less than P2, as shown in Fig. 2(f), due to the way the dislocation network meets the surface. The number of metastable GB states on a potential energy surface like this, with respect to the position of the GB dislocation network, increases with particle size, as seen by comparing Fig. 2(f) (5 nm) and Fig. 2(g) (20 nm). Therefore, the number of metastable configurations for GB dislocation networks appears to decrease with decreasing particle size which suggests the probability of finding a stable structure with $\Delta\theta \neq 0$ also decreases with particle size. This is consistent with the observations of Chan and Balluffi which showed nanoparticles align themselves with a bulk single crystal substrate [40-42]; however, full alignment was not observed in numerous experiments performed with micrometer-sized particles [43,44].

The size-dependent transition from the classical, continuous Read-Shockley behavior to a quantized regime with discrete states is also evident in [111]-twist GBs, as shown in Fig. 2(h).

Notably, there are no NTBs with twist misorientations in the range of 52º<θ<68º, implying particles with misorientations within this range of angles will rotate to remove elastic strain and transform to TBs. This discrete hopping toward local minima on the potential energy surface may explain the formation of the TB between particles A and B in Figs. 1(c)-1(e) and the abundance of TBs after 30 ns in Fig. 1(b); depending on the GB plane, the probability of a NTB disintegrating to form a metastable TB may be higher than the probability to fully disintegrate to Δθ = 0. However, as discussed above, junctions formed by two or more twin boundaries and a high angle boundary (Figs. S10, S11) or by five TBs, were found to pin the GBs and hinder single crystal formation. Thus, the two-particle GB energies described by these calculations provide only an idealized version of the energy landscape associated with GB in real systems which will depend on the local strain environment, nearby surface structure, and the presence of nearby extended defects such as additional dislocations in the bulk and GBs.

Grain evolution after nanoparticle attachment was explored experimentally by annealing small clusters of single crystal gold nanoparticles with diameters of ~20 nm at ~500 ºC in vacuuo in the transmission electron microscope (TEM) and capturing images at low magnification to minimize the effect of the electron beam. Clusters of misoriented, single crystal particles (Figs. S12 and S13) were intentionally prepared to capture GB evolution after agglomeration without the influence of preexisting extended defects. Two notable observations were made which are consistent with the GB disintegration process seen in the MD simulations presented above. First, a more compact, nearly single crystal aggregate was formed after annealing an agglomerate consisting of 9 misoriented particles for roughly 2 hours (shown in Fig. 3(a)). This validates that the thermal barriers to form a single crystal aggregate, even with 20 nm particles, are relatively small. Second, the GB removal process in Figs. 3(b) and (c) was accompanied by dislocation slip. A single grain prior to heating was identified by dark field imaging as shown in Fig. 3(c) and highlighted in blue in Fig. 3(b). After heating, the boundary of the highlighted grain extended to the edge of the adjacent grain and a defect was visible across the coalesced region (dark streak in Fig. 3(b) and light streak in Fig. 3(c)). High resolution imaging confirmed that this defect was a twin as shown on the right side of Fig. 1(c). Additional examples of particle cluster evolution during annealing are shown in Fig. S14.

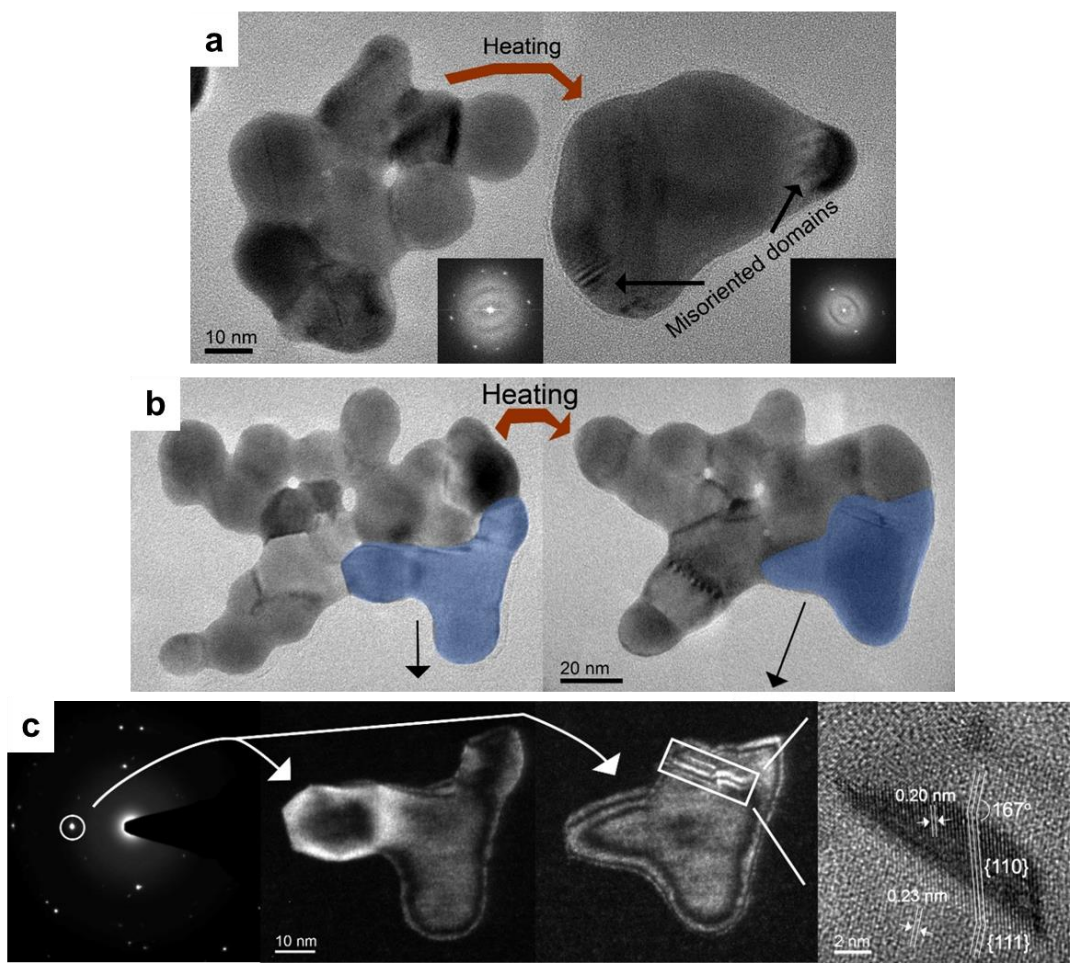

**Figure 3**. (a) Bright field TEM image showing Imperfect OA (right hand side) after annealing 9 misoriented, attached, single crystal particles (left hand side) with diameters of ~20 nm at ~500 C. (b) Bright field TEM images showing grain evolution in agglomerated, single crystal

nanoparticles after annealing at ~500 C for 1 hr. (c) Left hand side shows electron diffraction pattern of annealed agglomerate in (b); middle subfigures show dark field images, corresponding to images in (b) using circled diffraction spot on left hand side; right hand side shows high-resolution TEM image of twin protruding from GB after grain rotation.

It is clear from both the MD simulations and TEM results that freely sintered nanoparticle clusters readily evolve to form single crystal aggregates. The nudged elastic band calculations and potential energy surfaces discussed above indicate this is due to a dislocation slip processes (ii) as a result of a GB disintegration because:

(1) small barriers exist to move GB dislocations to the surface,

(2) there are fewer dislocations to move to the surface at the nanoscale,

(3) and only a discrete number of GB dislocation configurations are metastable which decreases the probability of observing GBs with $\Delta\theta \neq 0$.

A schematic is presented in Fig. 4 which aims to capture the Imperfect OA pathway under dry, free sintering conditions. This follows the style of the classical Swinkels-Ashby mechanism map for sintering in coarse-grained systems [45] and applies to the case of unpinned particles of a fixed size. An additional axis for particle size could be included to describe the transition from "GB disintegration" conditions toward those in which GB migration occurs (micrometer-sized particles). As shown by the arrows and cartoon in Fig. 4, local atomic shuffling occurs during attachment when the ratio of the neck to particle radius is small (light blue region). Once a stable GB dislocation configuration is formed, dislocation-mediated rotation occurs due to the glide GB dislocations to the surface (dark blue region). Finally, surface diffusion drives additional densification and shape evolution (yellow region). At high homologous temperatures, it is expected that additional mechanisms may become active such as surface pre-melting and GB migration.

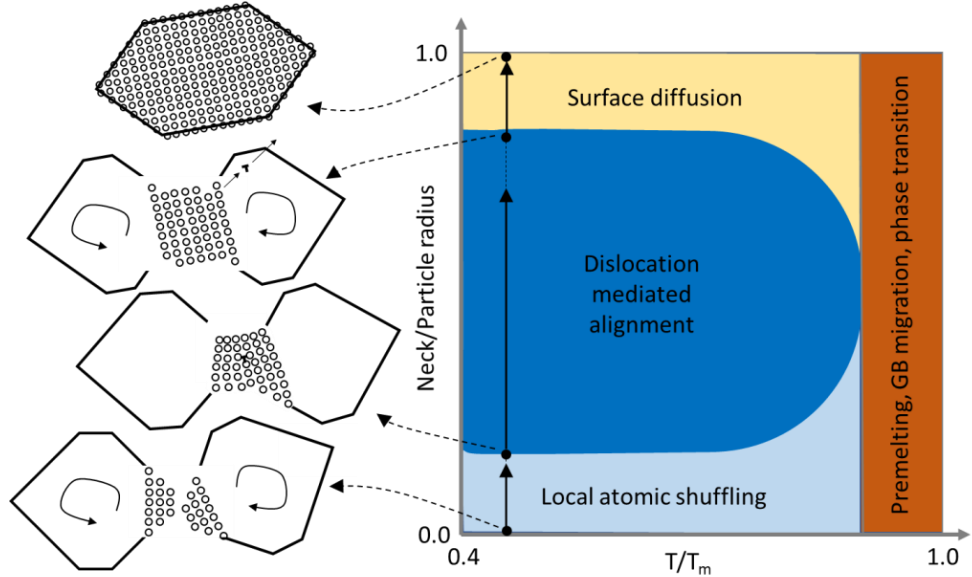

**Figure 4**. Proposed mechanism map in state-space of temperature and neck radius. Captures fundamental changes in coalescence mechanisms from local atomic shuffling during attachment to dislocation-mediated after attachment, underpinned by the quantized nature of GBs, to surface diffusion-controlled densification after sufficient densification.

The proposed concept of GB disintegration GB quantization motivate future improvements to phenomenological modes for grain rotation during nanoparticle sintering and crystal growth. Further, these results raise new questions about the behavior of GBs in bulk NC materials. For example, additional topological restrictions are placed upon GB networks due to this quantization which may lead to high internal elastic strain or disclinations [45]. To experimentally probe the discrete phenomena presented here, the population density of specific grain boundary misorientations in systems with hundreds or thousands of GBs during and after thermal equilibration must be measured. Alternatively, the rotation of individual particles during thermally driven aggregation must be probed on short time and length scales to capture thermal processes with low kinetic barriers. The latter approach, however, is challenging due to the influence of substrate or liquid media in in situ TEM and electron probe-particle interactions. Ultimately, the discretized nature of nanoscale GBs further provides a pathway to engineer extended nanostructures, colloidal crystals, or bulk nanocrystalline materials with tailored defect structures and electronic states.

**Materials and Methods**

Gold Nanoparticle synthesis

Twin-free, single crystal Gold nanoparticles were synthesized using a procedure published by Grzelczak et al.[47]. The particles were grown in two steps, seed nucleation and overgrowth, as described below.

*Seed nucleation*. In a 20 mL scintillation vial, 1.2 mL of milli-Q water was mixed with 31 μL of 10 mM HAuCl4. To this solution, 202 μL of 25 wt. % cetyltrimethylammonium chloride (CTAC) was added. Under rigorous stirring with a magnetic stir bar, 75 μL of 10 mM NaBH4 was added to the solution of water, HAuCl4, and CTAC. The resulting solution was then stirred for 2 hours at 27 °C.

*Overgrowth*. The gold nanoparticle seeds were subsequently overgrown to a diameter of 15-20 nm. The overgrowth solution was prepared in a separate 20 mL scintillation vial which consisted of 12.8 mL of milli-Q water, 1.3 mL of 25 wt. % CTAC, 0.25 mL of 10 mM HAuCl4, and 0.066 g of ascorbic acid. The solution was stirred using a magnetic stir bar. Under rigorous stirring, 27.5 µL of the gold nanoparticle seed solution was added to the overgrowth solution. The stir bar was then removed, and the solution was left to sit for 30 minutes. After 30 minutes, the gold nanoparticle solution was washed twice by centrifuging at 14000 rpm for 10 minutes and replacing the supernatant with water. The hydrogen tetrachloroaurate(III), ACS, 99.99 % (metals basis), sodium borohydride (99 %, from Sigma Aldrich), ascorbic acid (99 %, from J. T. Baker), and cetyltrimethylammonium chloride (CTAC) were used as-received. 18.2 M milli-Q water was used throughout the syntheses as needed. An Eppendorf 5415C centrifuge was used to remove the surfactant from the nanoparticle solution during the washing step and to concentrate the nanoparticle solution for TEM grid preparation, described below.

Transmission electron microscopy grid preparation

The synthesized nanoparticles were concentrated by centrifuging 90 µL of the final solution from the steps above at 14000 rpm for 10 minutes and removing 80 µL of the supernatant. From the resulting, concentrated gold nanoparticle solution, 10 µL was placed on a 100 nm thick amorphous SiNx transmission electron microscopy grid manufactured by Norcada Inc. and left to dry overnight. The grid was then plasma cleaned for 15 seconds using an Oxford Air Plasma Cleaner to remove ligands (CTAC) and environmental carbon contamination from the surface. The TEM images in Fig. S12 (A) and (B) show as-prepared particles (without plasma cleaning). It is clear from these images that the synthesized particles were single crystalline. Figure S12 (C) shows a representative image of the particle morphology in the sample that underwent plasma cleaning for 15 seconds. Note these particles were partially sintered; however they remained misaligned, presenting an ideal condition for the study. Figure S12 (D) shows an energy filtered TEM (EFTEM) image of the area in Fig. S12 (C) centered at the carbon K-edge. A higher concentration of carbon was observed over the area with the nanoparticles. Therefore, all the surfactant was not removed. The ratio of the carbon to gold energy dispersive X-ray spectroscopy peaks, however, was 1.4 in the examined area of the cleaned sample (Fig. S12 C) compared to 2.2 for the as-prepared sample (Figs. S12 B), indicating some reduction.

Transmission electron microscopy experimental details

*Beam heating.* While imaging particles at high resolution, unintentional heating of the support membrane can occur which causes coalescence [48]. In the supplemental video, S3, direct beam heating was utilized to capture atomistic details of the coalescence process.

*Holder heating.* Two heating experiments were carried out. Images and selected area diffraction patterns of 4 separate nanoparticle clusters were acquired prior to heating. During the first heating experiment, the temperature of the holder was set to and held at ~450 °C for ~90 minutes. Images and diffraction patterns of the 4 particle clusters of interest were then acquired. During the second heating experiment, the temperature of the holder was set to and held at ~570 °C for ~60 minutes. The clusters were again examined using selected area diffraction (SAED), high resolution transmission electron microscopy (HRTEM), and dark field (DF) TEM. The nanoparticles sat roughly 13.5 μm from the edge of the SiNx membrane. All TEM images were acquired using an FEI Titan TEM operated at 300 kV. A Gatan double-tilt heating/straining holder was used to heat the samples and an FEI double-tilt holder was used for the beam heating experiments.

Details of molecular dynamics simulations involving many particles

Gold nanoparticles with varied sizes, drawn from a normal distribution of nearly spherical particles with a mean diameter of 5 nm and standard deviation of 1.5 nm, were constructed using the Wulff method. In this construction, the length from the center of the particle to the surface along the surface normal is inversely proportional to the surface energies which were drawn from Wang et al. for this study [49]. The particles were then individually thermalized by increasing their temperatures from 100 to 1100 K at a rate of 4 K/ps using NVT constraints (constant number of particles, constant volume and constant temperature) and a Nose-Hoover thermostat. Pseudo-random misorientations were then applied to the particles prior to placing them adjacent to each other at separation distances less than the cutoff distance of the embedded atom potential developed by Ackland et al., 4.9786 [50]. Aggregates consisting of 2, 10, 15, 20, 25, and 30 particles were then evolved, again using NVT constraints, for 30-40 ns using a 1 fs time-step. To remove thermal fluctuations for subsequent analysis, the resulting structures were minimized using a 1000 step steepest decent algorithm and visualized using OVITO [51]. The structures were further analyzed using the dislocation extraction algorithm and common neighbor analysis algorithms available in OVITO. All simulations were run using the Large-scale Atmoic/Molecular Massively Parallel Simulator (LAMMPS) [52].

In a recent publication, Rassoulinejad-Mousavi and Zhang found that the embedded-atom-method potential used for the gold simulations does not accurately predict the shear modulus [53]. Therefore, simulations were run using a potential developed by Grochola et al. [54], found to best fit experimental values for elastic stiffness constants, starting with the same initial conditions as those reported in the manuscript for the potential developed by Ackland et al. Simulations were also run with the well-established potential developed by Mishin et al. for copper [55]. There were some variations in the observed mechanisms for these different potentials. For example, at a simulation temperature of 1100 K, surface diffusion was significantly enhanced when the Grochola was used. This caused Ostwald ripening to occur when smaller particles contacted larger particles due to surface premelting. However, dislocated-mediated alignment, in which grain boundary dislocations moved to the surface, was observed at 800 K. Therefore, it is believed the high-temperature boundary in Fig. 4 at which dislocation-mediation processes give way to other mechanisms may be difficult to predict using MD; however, the salient observations were reproducible.

Calculations of grain boundary stability in nano-scale materials

To understand the stability of grain boundaries (GBs) in nanoparticles GB energies were calculated for different misorientations and potential energy surfaces with respect to lateral translations within the GB plane were generated for a few different twist misorientations. Grain boundary structures were generated by separating two nearly spherical nanoparticles with the same radius (R = 1.5, 2.5, 5, 10, or 20 nm) by a distance of 0.6R. Tilt or twist misorientations were applied by rotating one of the particles and all overlapping atoms in the system were deleted to generate an abrupt grain boundary. The resulting system was then minimized using the conjugate gradient method with a force tolerance of $10^{-8}$ eV/Å.

The stability of [010] tilt grain boundary dislocation networks was evaluated by performing NVT (constant number of atoms, volume and temperature) simulations at 20 K on the minimized structures for 100 ps. In trajectories with grain boundary dislocation glide events, structures at the beginning and end of the glide paths were used to locate the transition state and calculate the energy barrier associated with moving a grain boundary dislocation to the surface using the nudged elastic band minimum energy path calculation method. For [100] twist grain boundary energy calculations, no thermalization was performed; grain boundary energies were simply determined as a function of misorientation angle after minimization.

For a fixed misorientation, the atomic structure at the interface between two nanoparticles has many degrees of freedom (e.g. local atomic density, relative lateral displacement between the two nanoparticles). Minimizing the grain boundary energy with respect to local atomic density is a difficult proposition that requires global optimization within the grand-canonical ensemble. Therefore, only the degree of freedom with respect to lateral translations within the GB plane were considered. Potential energy surfaces with respect to lateral translations within the grain boundary plane were calculated to determine minimum energy structures at fixed [100] twist misorientation angles. We started with two rectangular blocks with a specified misorientation in contact such that the interface consisted of (100) planes. The lattice of one block was incrementally translated in the [010] and [001] directions, after which a two-particle structure was defined with respect to a fixed point to ensure the grain boundary perimeter was well aligned. The structure was then minimized using the conjugate gradient method with an energy tolerance of $10^{-8}$ eV/°A.

Calculations for thermally activated events on nanoparticle surfaces

The surface of a nanoparticle is a rich source and sink of defects. Such defects, although spatially localized, can initiate processes that can lead to dramatic changes in surface morphology. Thus, energy surface exploration techniques were used, along with molecular dynamics, to analyze the energy barriers and kinetics of thermally activated events that play an important role during nanoparticle coalescence. Since these calculations are computationally expensive, copper was used as a model system. Since these events happened frequently during copper and gold nanoparticle coalescence simulations, we believe the trends are generally valid for metals.

*Energy barrier for homogeneous nucleation in bulk.* Elastic-plastic transition through the nucleation of dislocations from surface or bulk sources have quite different statistical behavior from that of the Frank-Read type sources operative in the bulk of a material, or the self-organized criticality exhibited by dislocations [56,57]. To calculate the dislocation nucleation barrier in bulk, we used a supercell with lattice vectors parallel to [100], [010] and [001] directions and containing 108,000 atoms (108.45×108.45×108.45 Å). The energy barriers were calculated by using the transition state theory-based tool - climbing image nudged elastic band method [58]. The activation energies for homogeneous nucleation as a function of the resolved shear stress are shown in Fig. S8. The barrier is very high at stresses below 3 GPa and is below 0.5 eV only for stresses above 4.5 GPa. Since no external stresses were imposed on the nanoparticles during MD simulations, it is thus believed that dislocation nucleation cannot happen in the bulk of nanoparticles as this would require stresses close to the yield strength of Cu or Au.

*Energy barrier for shearing of surface islands.* A nearly spherical nanoparticle with 5 nm radius was used to calculate energy barriers for shearing small surface islands. The energy barriers were calculated by using a tool based on transition state theory named the climbing image nudged elastic band method [58]. Fig. S9 shows the energy barrier to shear a 30-atom island is 0.64 eV while the barrier to shear three successive layers from the top is close to 4 eV. The energy barrier to split a 12-atom island is 0.70 eV, little higher than the energy required to shear the whole island.

ASSOCIATED CONTENT

Supplementary Material

AUTHOR INFORMATION

**Corresponding Author**

*Selim Elahdj
Lawrence Livermore National Laboratory

Mail-stop L-470

7000 East Ave. Livermore, CA 94550

**Funding Sources**

This work was performed under the auspices of the U.S. Department of Energy by Lawrence Livermore National Laboratory under Contract DE-AC52-07NA27344. LLNL-JRNL-753683. Tracking # LDRD 15-ERD-057.

ABBREVIATIONS

OA, oriented attachment, MD, molecular dynamics; TEM, transmission electron microscopy; GB, grain boundary; NC, nanocrystalline; NTB, non-twin boundary; TB, twin boundary; NEB, nudged elastic band.

**Supplementary Material**

**Movie S1**

Movie showing molecular dynamics simulation of 10 initially misoriented gold nanoparticles during aggregation. Salient features are called out during movie which show dislocation-mediated alignment process of small particles on perimeter then slow disintegration of grain boundary in interior of aggregate. Simulation run using potential developed by Ackland et al.[50].

Additional plots of non-twin boundaries (NTBs) and twin boundaries (TBs) as a function of sintering are provided in Fig. S2. Similar trends observed in all simulations:

- The number of NTBs rapidly decreased during the early part of the simulation
- The number of NTBs then slowly decreased as more stable GBs were removed
- The number of TBs rapidly increased due to the formation of twins from surface shearing events and the evolution of NTBs to TBs
- The number of TBs then slowly decreased
- The final structures generally had GB junctions with multiple stable TBs and at least one stable NTB

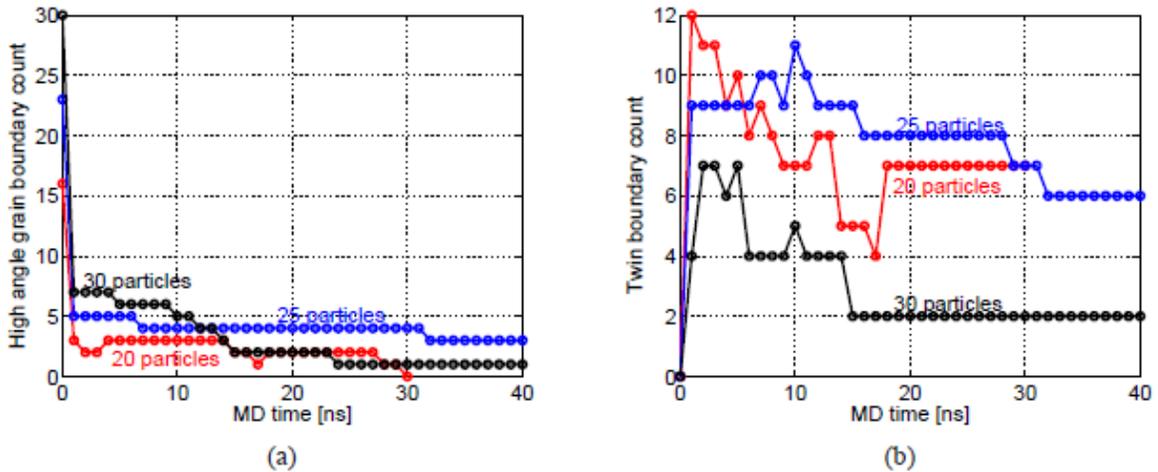

**Fig. S2.**

Plot of the number of non-twin grain boundaries and coherent twin boundaries with simulation time for simulations involving 20, 25, and 30 particles.

Supplementary figure S3 shows the attachment and GB disintegration process in greater detail for the particles marked A and B in Fig. 1A. As the {111} planes of the two particles bond together the smaller particle rotates relative to the larger particle. During this process, local atomic shuffling at the interface facilitates the formation of edge dislocations with Burgers vectors of a[111]. The edge dislocations then climbed as the GB continued to form. Finally, the top particle sheared which left a TB at the interface. This presumably occurred by the dissociation and glide of GB dislocations to the surface.

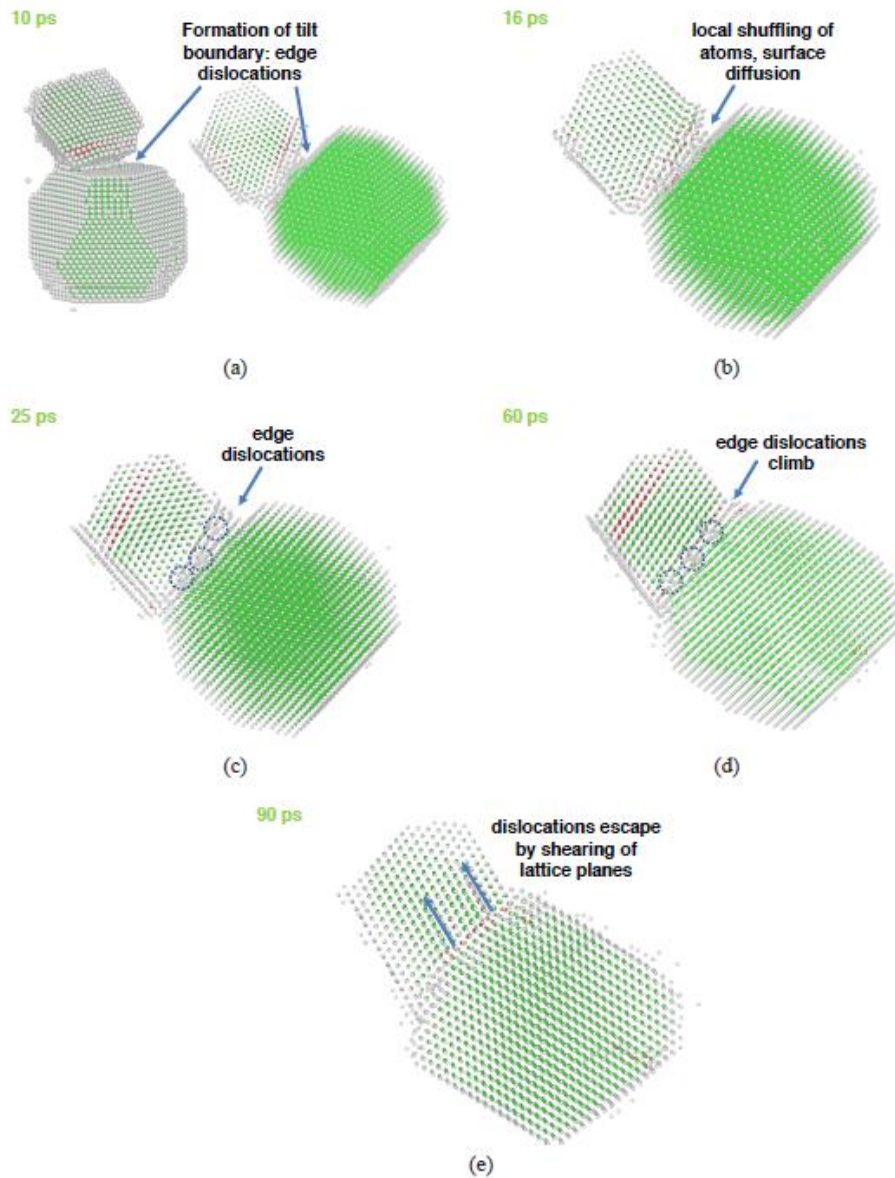

**Fig. S3.**
Intermediate structures during the coalescence of two nanoparticles marked A and B in Fig. 2A. Local atomic arrangement at the interface aided by surface diffusion led to the formation of a well-defined GB, shown in (c), consisting of three dislocations. Since lattice planes in the GB contained tens of atoms, diffusion and local atomic shuffling helped these dislocations climb, leading to the structure in (d). Subsequently, the dislocations glided through the top particle (particle B) and exited at the surface as shown in (e).

Beam heating of small, 5-20 nm particles was carried using citrate-coated particles provided by BBI Solutions as well as synthesized in-house. These particles were heavily twinned, which made it difficult to track grain evolution. However, two notable observations were made during these experiments. The first was the relative rotation between two particles after attachment which led to grain boundary disintegration. Particle rotation after attachment was observed previously, but in this case, the grain boundary was eliminated which necessitates a concurrent evolution of the grain boundary character.

In movie S4, the lattice planes at the interface of the two, initially misoriented gold nanoparticles became aligned within seconds. The particles rotated after neck formation, which was followed by neck growth, then shape evolution. The neck grew until its width was roughly equal to the diameter of the smaller of the two particles. While neck growth was completed within minutes, subsequent shape evolution continued for a few more minutes. These observations qualitatively illustrate the sequence and relative time scales of these different aggregation processes under beam-heating conditions. Specifically, the alignment of lattice planes occurs during neck growth followed by shape evolution which is thought to be rate-limited by surface diffusion in nanoparticle systems. Another example of particle alignment during neck growth is shown in movie S5. In this case, crystallographic alignment across the interface is easier to discern.

**Movie S4.**
Movie showing 2-particle coalescence during *in-situ* TEM electron-beam heating experiment. This observation highlights the relative timescales of the coalescence process where particle rotation occurs during attachment, followed by neck growth, then finally shape evolution.

**Movie S5.**

Movie showing 2-particle coalescence during *in-situ* TEM electron-beam heating experiment. This observation highlights particle rotation during neck growth. As the neck grows, evident from successive lines at interface, the particle on the right rotates, evident from successive traces of atomic planes in left-hand particle.

*Kinetics of surface shearing*

The energy barriers for shearing small surface islands, described in the methods section and shown in Fig. S6, are less than 1 eV which is significantly lower than barriers to nucleate dislocations in the bulk shown in Fig. S7. This suggests that such events can happen frequently even at room temperature, especially if strain is present near grain boundaries due to surface tension. For example, using the Arrhenius relationship $f = f_o \exp(E/k_bT)$, taking $f_o = 1 \times 10^{13}$ sec$^{-1}$ (rough oscillation frequency of surface atoms in potential well), T = 300 K, $k_B$ = 8.617×10$^{-5}$, E = 0.75 eV, it is found f ~ 2.5 sec$^{-1}$. If the barrier is reduced to 0.5 eV, the frequency of surface shearing is found to be ~4 x 10$^4$/sec. At higher temperatures (~1000 K), the frequency of such events increases to ~1x10$^9$ sec$^{-1}$, comparable to the timescale of molecular dynamics simulations. The calculations shown in Fig. S6 represent only a couple ways in which dislocations can nucleate from the surface of a nanoparticle. A movie showing only atoms with hcp coordination during a simulation of a 5 nm particle annealed at 1100 K is provided in movie S8. This visualization scheme was used as bulk atoms have fcc coordination while hcp-coordinated atoms only arise in the plane below sheared domains on the surface. It is clear from this video that the lifetime of stacking faults near the surface varied considerably for the Ackland MD potential; some remain for hundreds of picoseconds, while others disappear after less than ten picoseconds.

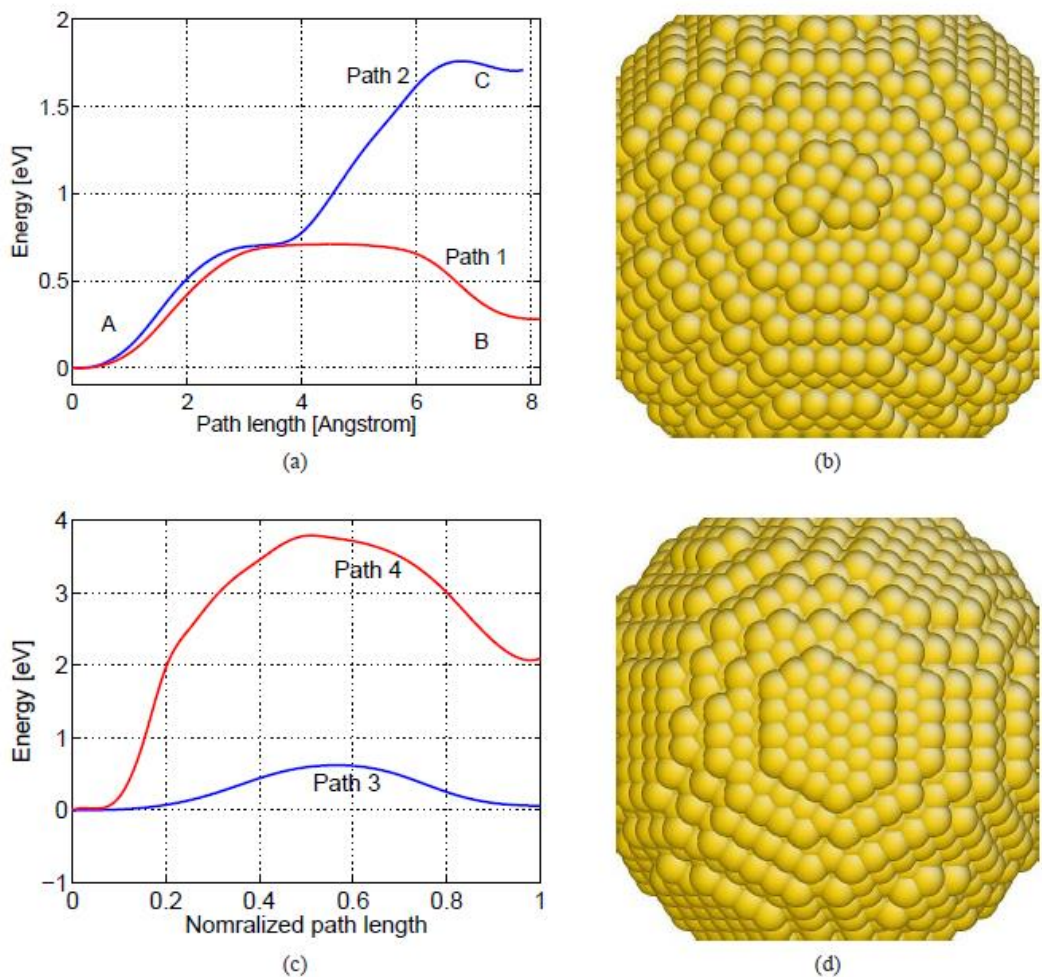

**Fig. S6.**

(a) shows the energy barrier to shear two rows of atoms (containing 5 atoms in total) in a 12 atom island (marked Path 1) is 0.70 eV. The blue curve (marked Path 2) shows the energy barrier to further shear the 5 atoms further away from the remaining 7 atoms is 1.03 eV. (b) shows the saddle point (i.e. the transition state) along Path 1. (c) shows the energy barrier to shear one island (marked Path 3) and top three layers (marked Path 4) in the nanoparticle shown in (d).

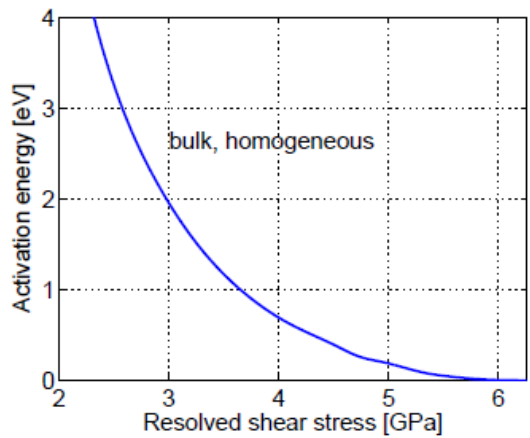

**Fig. S7.**
Shown here is the activation energy for homogeneous nucleation of a dislocation in bulk Cu as a function of the shear stress resolved on a (111) plane and along [112] direction.

**Movie S8.**
Movie of molecular dynamics simulations over 1 nanosecond showing a single particle annealed at 1100 K. Only atoms with hcp coordination are shown corresponding to layers under sheared surface domains and each frame in the video corresponds to a snapshot after 100 femtoseconds. It is clear from this video that the surface of the spherical particles here, under the chosen MD potential, are very active providing a rich source and sink of particle dislocations across the neck-groove in agglomerated particles.

*Grain boundary potential energy surfaces*

The grain boundary energy profiles shown in Fig. 2B and C were calculated at fixed y and z, y = z = 0, with x being the direction normal to the grain boundary. In the case of 10 nm particles, this position corresponded to a grain boundary energy minimum (Fig. 2G). However, this position did not correspond to a minimum in the case of 5 nm particles as shown in Fig. 2F.

Many GB energy minima with respect to lateral translations within the GB plane are evident in Fig. 2F and G. These minima are associated with low energy dislocation network configurations. The potential energy surfaces differed significantly for systems with the same misorientation but different particle size. This indicates that particle size influences the way dislocation networks "tile" the grain boundary. The potential energy surfaces for systems with the same particle size but different misorientations had the same overall pattern, or symmetry, while the number of minima increased for higher misorientation angles. Thus, particle misorientation determines the dislocation separation while translations in the grain boundary plane determine how the dislocation network sits on the grain boundary. The latter is expected to be more significant for smaller particles for which a larger fraction of the grain boundary is near the grain boundary perimeter.

*Model to describe discrete grain boundary energy levels*

The discrete nature of grain boundary energy profiles can also be explained by considering the GB schematic below. Imagine the distance between dislocations corresponds to that determined by the Read-Shockley model for a given misorientation (blue lines in figure below).

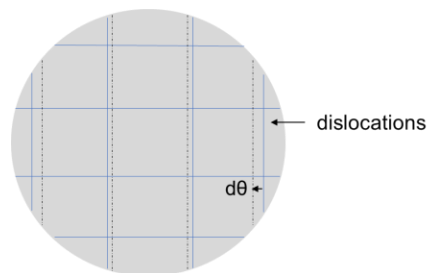

If the misorientation increases, the separation between dislocation segments (in an infinite boundary) should decrease as shown by the dotted black lines. Although the total dislocation length, and therefore grain boundary energy, will increase slightly in this case due to the curvature of the boundary perimeter, it will increase much more abruptly when the dislocation separation is such that new additional dislocation line segments are introduced at the edges of the GB. It should be noted this simple model does not account for dislocation interactions which will play a more dominant role as the misorientation angle increases and the separation decreases. However, it highlights one cause of the observed discretization. When particle diameters are very small (say, 2.5 nm), the number of dislocations at the GB, even at large misorientation angles, may only equal one or two. In calculations of GB energy with twist misorientation, it was found that particles rotated during minimization so that there is only one stable dislocation array (or GB structure) corresponding to the addition of each new dislocation line segment. This effectively eliminates the continuity in GB energy with misorientation as results in the formation of one stable misorientation "state" for each misorientation "band", or range of angles.

If translations within the GB plane are also considered, the number of states for a given misorientation angle would increase. Changes in lateral translations of particles are equivalent to changes in the way the dislocation arrays "tile" the GB. Therefore, for a given misorientation, there should be one (or more) dislocation configurations which correspond to minimum energy states (dictated by the way the dislocations intersect the GB perimeter).

The total energy of dislocations present at the interface between two nanoparticles consists of four different contributions:

>    *(i)* self-energy of all dislocations in the grain boundary,
>
>    *(ii)* energy due to interaction between dislocations in the grain boundary,
>
>    *(iii)* energy due to interaction between dislocations and their surface images, and
>
>    *(iv)* energy of dislocation junctions.

The self-energy of a straight dislocation line (ignoring the end effects) is $E_s = 0.5\mu b^2 L$, where $\mu$ is the shear modulus on the glide plane and along the Burgers vector direction, b is the magnitude of the Burgers vector of the dislocation and L is the length of the dislocation. The interaction energy between two dislocations is $E_i = \mu b^2 L \log(r/r_o)$. Here, $r \ll L$ is the separation between dislocations and $r_o$ is the core-cutoff radius. The interaction between two parallel dislocations with same Burgers vector is repulsive while the interaction between a dislocation and its surface image is attractive. Within the Read-Shockley theory, the separation between parallel dislocations in a grain boundary with twist misorientation $\theta$ is $d_\theta = b/2 \sin(\theta/2)$. For small $\theta$, $2\sin(\theta/2) \sim \theta$ and $d_\theta = b/\theta$. A calculation for the total energy of a GB dislocation within a GB of radius R, considering only (*i*) and assuming a dislocation separation predicted by Read-Shockley about the GB center, is shown in Fig. S9. This is accurate for small angles when $r/r_o$ is much greater than 1 (i.e. $r/r_o \gg 1$) as $\log(r/r_o)$ changes very slowly with changes in $\theta$ and can be approximated to be a constant, eliminating the influence of (*ii*) and (*iii*). Figure S9 shows that the self-energy of dislocations vs misorientation agrees well with the simulation results for low angle grain boundaries (< 15 degrees) when the separation between dislocations is large. For high misorientation angles, dislocation separation is small and which can lead to the formation of alternative high angle GB structures, such as coincident site lattices. This reduces the overall energy of the GB (as seen in Fig. S9 at larger misorientations), however the discrete nature of misorientation states was still observed for these more complex GB structures. It should be emphasized again here that the GB energy for the MD calculations is plotted against the <u>starting</u> misorientation between particles in Fig. S9. After minimization, the relative misorientation was always different (e.g. going to 0º for all misorientations below 5º). Therefore, a more accurate representation of this data would be discrete lines similar to what was plotted in Figs. 2 B, C, and H.

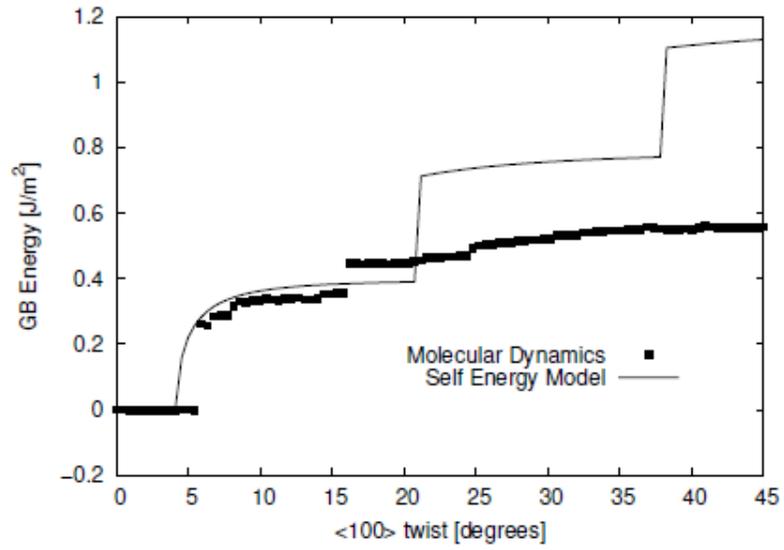

**Fig. S9**

The total self-energy of dislocations in a model symmetric tilt grain boundary in a nanoparticle shows discrete jumps as the misorientation changes.

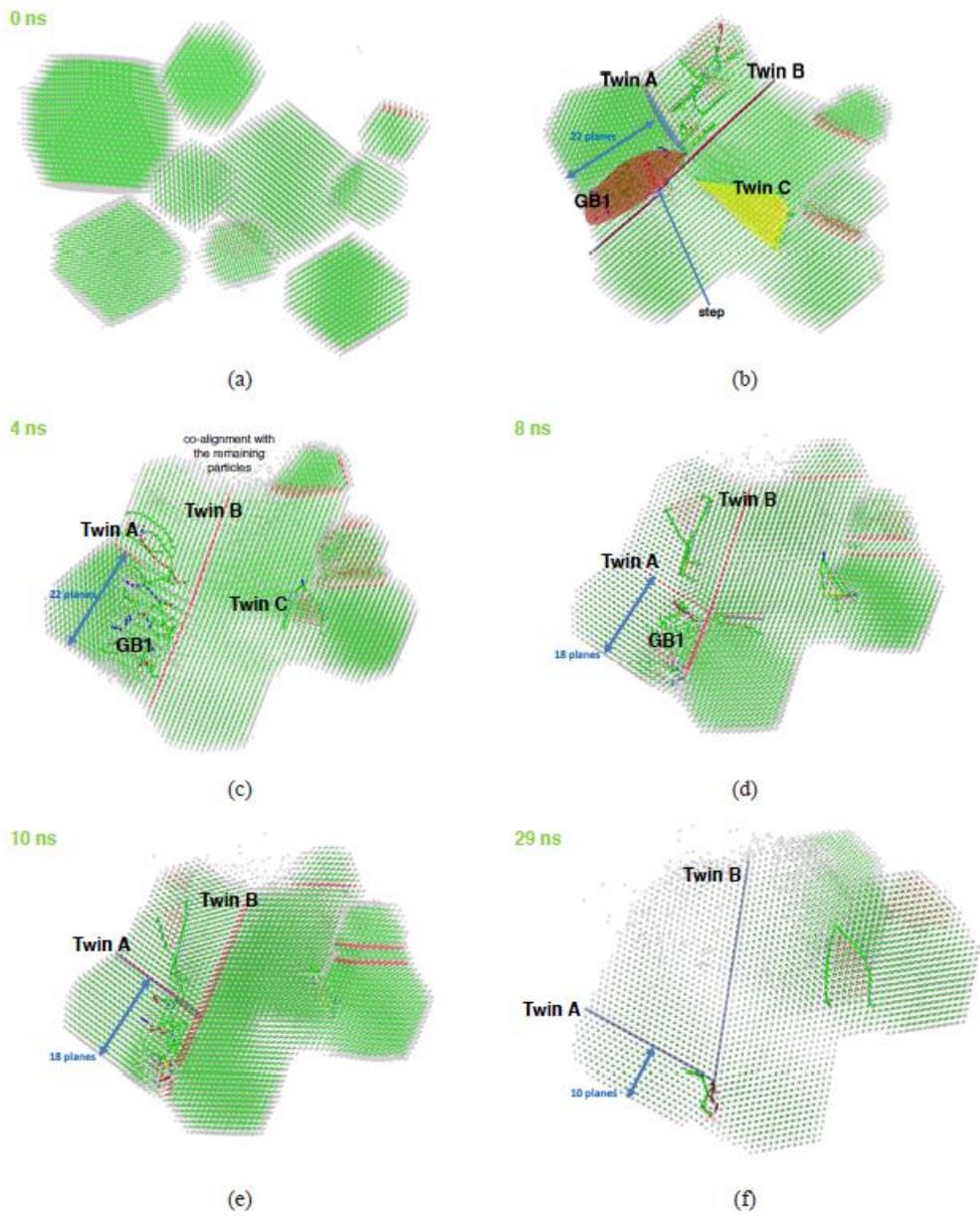

**Fig. S10**
Grain boundary junctions were observed to slow down the kinetics of single crystal aggregate formation. The coalescence of 10 Wulff-shaped nanoparticles, from a simulation with 25 gold nanoparticles in total, is shown in (a). The structure in (b) contains a complex junction involving three twin boundaries and a grain boundary. The twin boundary marked "Twin A" gradually migrated towards the surface in (c)-(f). During this time, the boundary marked "Twin C" shrunk in size and vanished.

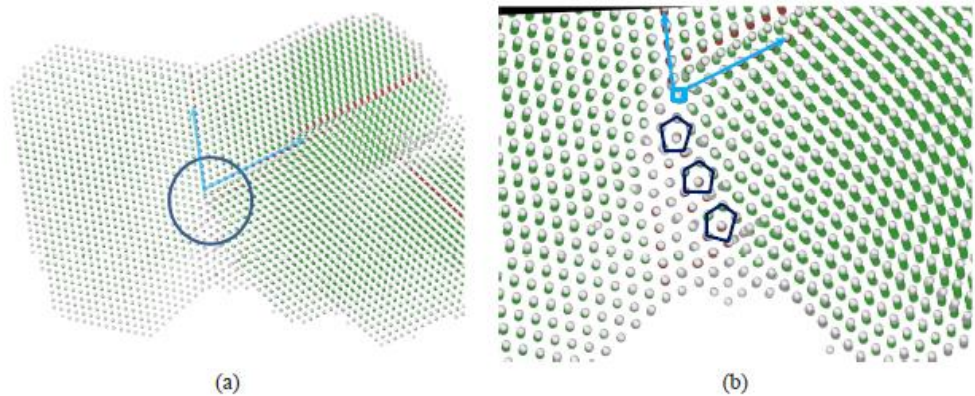

**Fig. S11**
Grain boundary junctions were observed to slow down the kinetics of single crystal aggregate formation. Shown here is the structural evolution of a high angle grain boundary between a triple junction and the surface in a 20 particle coalescence simulation. The grain boundary structure in (b) remains metastable for about 15 nanoseconds thus preventing single crystal formation as seen in Fig. 1.

The 4 clusters monitored during the heating experiments are labeled in Fig. S13(A). Additional low magnification images of the clusters after the first and second heating cycles are provided in Figs. S13 (B), (C), respectively. Higher magnification images of the two clusters not discussed in the manuscript, clusters 1 and 2, are shown in Fig. S14. The clusters are shown both before annealing (left) and after both annealing cycles (right).

The two most remarkable results from the holder heating experiments were the formation of a nearly single crystal agglomerate (cluster 3) and the presence of a remnant twin after a particle rotated to align itself with a neighboring grain. These two observations are highlighted in Fig. 1 of the manuscript.

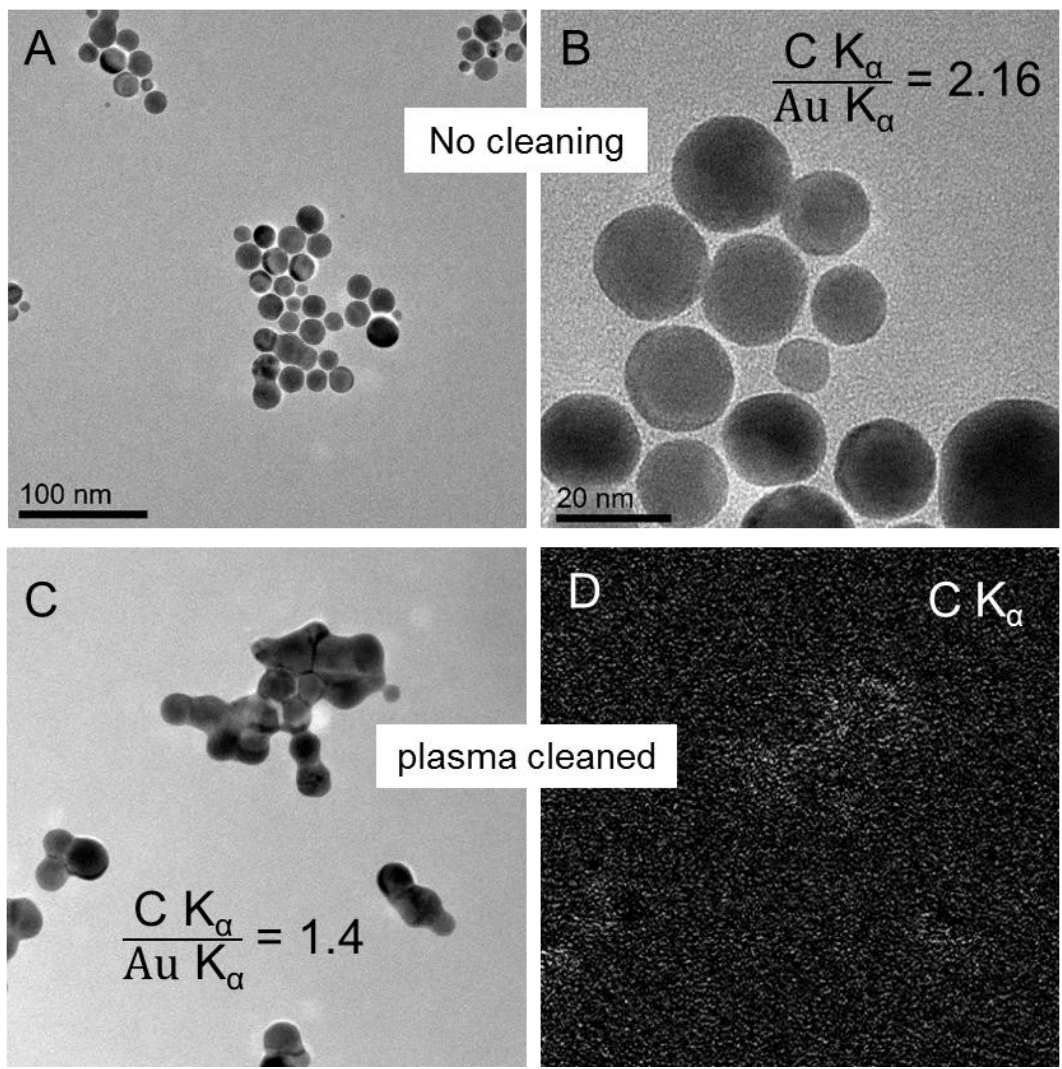

**Fig. S12.**
Bright field TEM images of as-synthesized nanoparticles after drop-casting (A and B), and after drop casting followed by oxygen plasma cleaning for 15 seconds (C). Evidence of residual carbon on particle clusters evident in energy filtered TEM image tuned to the carbon K-edge in (D).

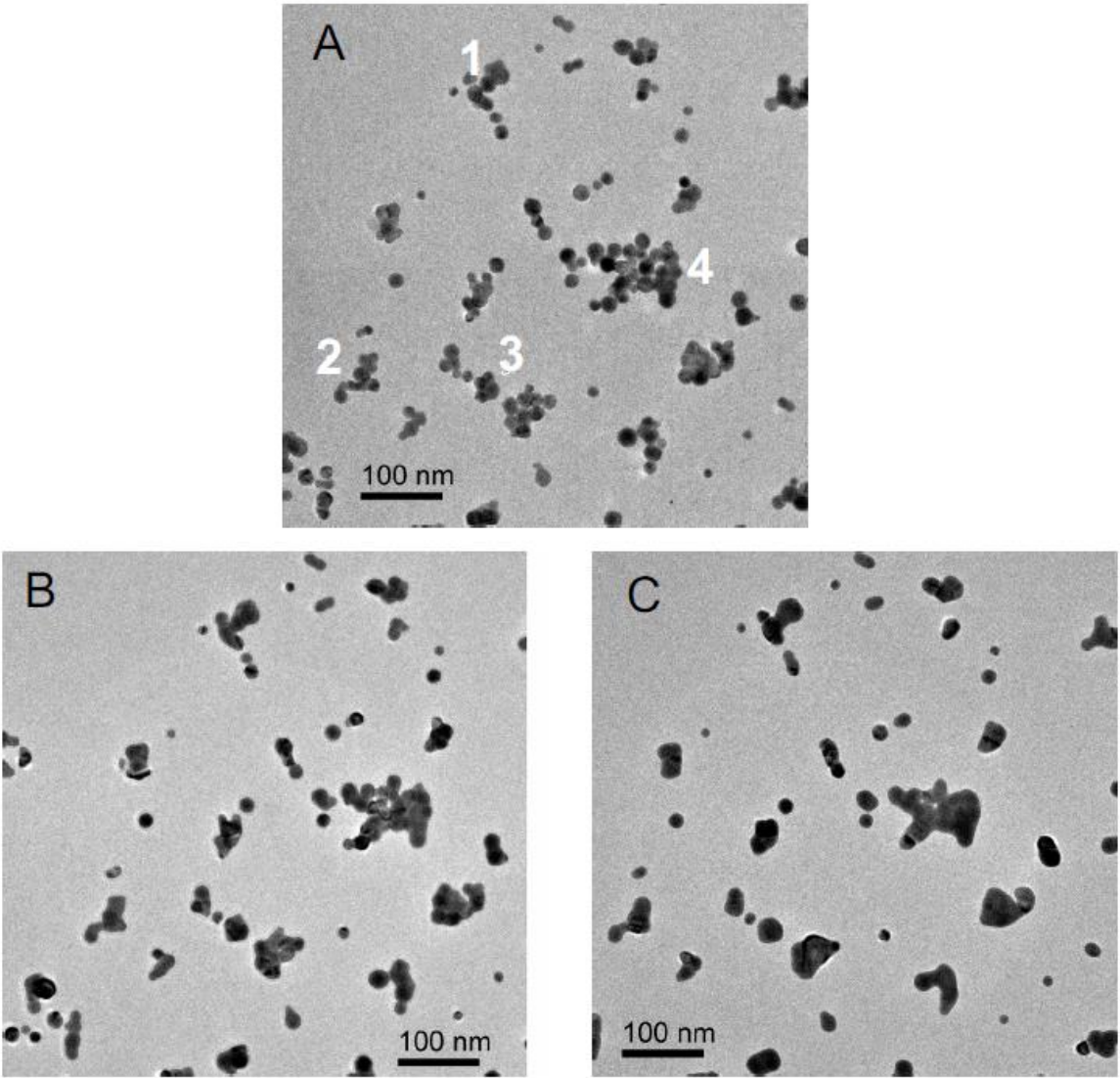

**Fig. S13.**
Particle clusters observed during two annealing cycles. Nanoparticles in clusters, labeled 1-4 in (A), were in contact but misoriented prior to annealing. Nanoparticles in each cluster underwent densification and alignment during ∼1.5 hour anneal cycle at 450 ◦C (B) and during ∼1 hour anneal cycle at 570 ◦C (C). The actual temperature experienced by the clusters depends on its proximity to the SiNx grid. So, it is possible that some of the clusters were at a much lower physical temperature and hence received only a small fraction of the supplied thermal energy.

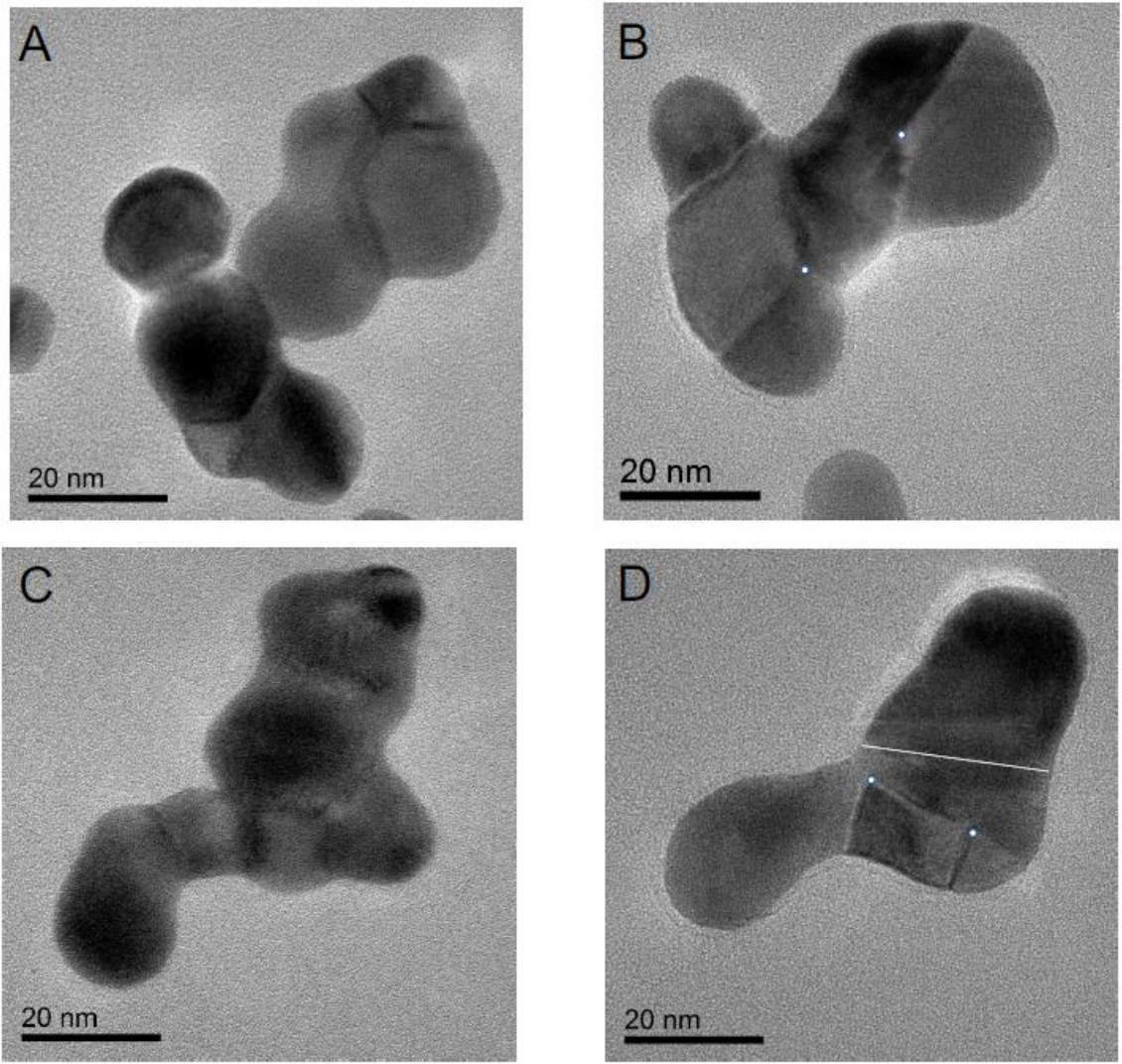

**Fig. S14.**
Particle cluster 1 (top) and 2 (bottom), as labeled in Fig. S13, before (left) and after (right) heating at ~450 ◦C for ~1.5 hours and ~570 ◦C for ~1 hour. The actual temperature experienced by the clusters depends on its proximity to the SiN$_x$ grid. So, it is possible that some of the clusters were at a much lower physical temperature and hence received only a small fraction of the total supplied thermal energy.